\documentclass[twocolumn,showpacs,preprintnumbers,amsmath,amssymb,prb]{revtex4}

\usepackage{graphicx}
\usepackage{dcolumn}
\usepackage{bm}

\begin{document}

\title{Spectrum of low energy excitations in the vortex state: \\
comparison of Doppler shift method to quasiclassical approach}

\author{T.~Dahm, S.~Graser, C.~Iniotakis, and N.~Schopohl}

\affiliation{Institut f\"ur Theoretische Physik, 
         Universit\"at T\"ubingen, 
         Auf der Morgenstelle 14, D-72076 T\"ubingen, 
         Germany}

\date{\today}

\begin{abstract}
We present a detailed comparison of numerical solutions of the 
quasiclassical Eilenberger equations with several approximation
schemes for the density of states of $s$- and $d$-wave 
superconductors in the vortex state, which have been used recently.
In particular, we critically examine the use of the Doppler shift 
method, which has been claimed to give good results for
$d$-wave superconductors. Studying the single vortex case we
show that there are important contributions coming from
core states, which extend far from the vortex cores into the
nodal directions and are not 
present in the Doppler shift method, but significantly affect 
the density of states at low energies. This leads to sizeable
corrections to Volovik's law, which we expect to be sensitive
to impurity scattering. For a vortex lattice we also show
comparisons with the method due to Brandt, Pesch, and Tewordt
and an approximate analytical method, generalizing a method due
to Pesch. These are high field approximations strictly valid
close to the upper critical field $B_{c2}$. At low energies the
approximate analytical method turns out to give impressively
good results over a broad field range and we recommend the
use of this method for studies of the vortex state at not 
too low magnetic fields.
\end{abstract}

\pacs{74.20.Rp, 74.60.-w, 74.25.Bt}

\maketitle

\section{Introduction}
\label{secI}

There is now a growing number of candidate systems for unconventional
or strongly anisotropic superconductivity. Besides the high-$T_c$
cuprates, which are believed to be $d$-wave superconductors, there
are indications for unconventional superconductivity in Sr$_2$RuO$_4$,
\cite{Sigrist} UPt$_3$, \cite{Heffner} an organic 
superconductor, \cite{Lee} the 
$\kappa$-(ET)$_2$ salts \cite{Carrington} and a possible unconventional,
ferromagnetically driven superconductivity is discussed in the
recently found superconducting ferromagnets ZrZn$_2$, URhGe and UGe$_2$.
\cite{Kirkpatrick}
Also, in MgB$_2$ phonon mediated, strongly anisotropic \cite{Haas,Posa} 
or two-gap scenarios \cite{Liu} have been proposed. In such systems 
the strong momentum
dependence of the superconducting order parameter can lead to 
interesting new behavior, especially if there are gap nodes
present at the Fermi surface. In particular, the low-energy
excitations in the vortex state of such systems are expected to
display unconventional behavior, as has been studied recently.
\cite{Volovik1,Schopohl1,MacDonald,SimonLee,Morita,Knapp,Schachinger,Vekhter,MolerPRB,FranzPRB,WonMaki}
In these studies the so-called Doppler shift method has been used
frequently. \cite{Schachinger,Vekhter,MolerPRB,FranzPRB,WonMaki}
In this method, which dates back to the early days of the theoretical 
study of type~II superconductors, \cite{Maki} the quasiparticle 
excitation energy
is approximated by taking into account only the Doppler shift
due to the local supercurrent flow. While this method neglects
vortex core scattering and certainly is a bad approximation for
conventional $s$-wave superconductors, Volovik
has argued that in the case of superconductors having gap nodes, 
the low energy
excitations are dominated by the extended quasiparticle states
outside the vortex core and the Doppler shift method thus should 
give much better results, at least for high-$\kappa$ superconductors
at sufficiently low external magnetic fields. \cite{Volovik1} 
Based on the Doppler shift
method, Volovik predicted that the density of states at the Fermi level
in the vortex state of a superconductor with line nodes should vary as the
square root of the magnetic field, instead of the linear variation
expected for a conventional superconductor. Indeed, low-temperature
specific heat measurements on high-$T_c$ cuprates have been shown to
be consistent with such a field dependence.
\cite{MolerPRB,MolerPRL,Revaz,Wright,YWang}

However, the Doppler shift method is not a rigorous quasiclassical
approximation in the sense of an expansion in terms of $(k_F \xi)^{-1}$,
where $k_F$ is the Fermi momentum and $\xi$ the coherence length.
Instead, the appropriate method is the solution of the Eilenberger
equations for the quasiclassical propagator. 
\cite{Eilenberger1,LarkinOv1,SereneRainer}
Within this approach,
the contribution of vortex core states and vortex core scattering 
is included. Of course, the Eilenberger approach requires the 
solution of a set of transport equations, which is more involved 
than the Doppler shift calculations.

In the present work we want to present detailed comparisons of
Doppler shift calculations with solutions of the Eilenberger
equations for $s$- and $d$-wave superconductors, in order to
clarify where the Doppler shift method is a good approximation
and where it fails to give quantitative agreement with the
fully quasiclassical solution of the Eilenberger equations.
We are also going to discuss two other approximate methods,
which are expected to be good at magnetic fields close to
the upper critical field $B_{c2}$. Indeed, in this field range
it turns out that analytical results for the density of states
averaged over a unit cell of the vortex lattice can be obtained,
which are superior in accuracy than the Doppler shift results.
A numerical solution of the
Eilenberger equations is simplified considerably due to a recently 
found parametrization, which transforms the Eilenberger equations
into scalar differential equations of the Riccati type.

In the next section we want to briefly present the Eilenberger
approach and its mapping to scalar Riccati equations. We will
outline, how the Doppler shift method can be obtained from the
Riccati equations. Section \ref{secphase} is devoted to the study
of the local density of states for the single vortex, while in 
section \ref{seclatt} we are going to present comparisons of the 
density of states for the vortex lattice.

\section{Eilenberger approach}
\label{seceil}

For a layered system with cylindrical Fermi surface and a 
spin-singlet superconducting order parameter the Eilenberger
equation for the normal and anomalous components $g$ and $f$ of the
quasiclassical propagator reads \cite{Eilenberger1,LarkinOv1}

\begin{eqnarray}
\lefteqn{
\left[ 2 \left( i \epsilon_n + \frac{e}{c} \vec{v}_F \cdot \vec{A}
\right) + i \hbar \vec{v}_F \cdot \vec{\nabla} \right] f(\vec{r}, \Theta, 
i \epsilon_n) = } & & \nonumber \\ 
& & \hspace*{3.3cm} 
2 i g(\vec{r}, \Theta,i \epsilon_n) \Delta_0 (\vec{r},\Theta)
\label{eq1}
\end{eqnarray}
Here, $\vec{v}_F$ is the Fermi velocity pointing into the direction
$\Theta$, $\vec{A}$ is the vector potential due to the internal
magnetic field within the system, $\epsilon_n$ are the fermionic 
Matsubara frequencies, and
$\Delta_0 (\vec{r},\Theta)$ denotes the spacially varying and
momentum dependent order parameter. Eq. (\ref{eq1}) has to be
supplemented by a normalization condition, which 
in our case reads \cite{Eilenberger1}
\begin{equation}
\left[ g(\vec{r}, \Theta,i \epsilon_n) \right]^2 + f(\vec{r}, \Theta, 
i \epsilon_n) f^*(\vec{r}, \Theta + \pi, i \epsilon_n) = 1
\label{eq2}
\end{equation}
The normalized local density of states $N(\vec{r}, \epsilon)$ is 
obtained from $g$ after an analytic continuation to real frequencies
and an angular average over the Fermi surface:
\begin{equation}
N(\vec{r}, \epsilon) = \frac{1}{2\pi} \int_0^{2 \pi} d \Theta \; \mbox{Re} 
\left\{
g(\vec{r}, \Theta,i \epsilon_n \rightarrow \epsilon + i 0^+) \right\}
\label{eq2b}
\end{equation}
It has been shown in Refs. \onlinecite{Schopohl1,Riccati} that
Eqs. (\ref{eq1}) and (\ref{eq2}) can be mapped onto scalar
Riccati equations along real space trajectories $\vec{r}(x)$
running parallel to the direction of the Fermi velocity, if
one introduces two scalar complex quantities $a(x)$ and $b(x)$
\begin{equation}
g(\vec{r}(x)) =  \frac{1-a(x) b(x)}{1+a(x) b(x)} \qquad
f(\vec{r}(x)) =  \frac{2 i a(x)}{1+a(x) b(x)}
\label{eq3}
\end{equation}
where $a(x)$ and $b(x)$ obey the following Riccati equations
\begin{eqnarray}
\hbar v_F \frac{\partial}{\partial x} a(x) + \left[ 2 \tilde{\epsilon}_n(x) +
\Delta^\dagger(x) a(x) \right] a(x) - \Delta(x) &=& 0 
\nonumber \\
\label{eq4a} \\ 
\hbar v_F \frac{\partial}{\partial x} b(x) - \left[ 2 \tilde{\epsilon}_n(x) +
\Delta(x) b(x) \right] b(x) + \Delta^\dagger(x) &=& 0  
\nonumber \\
\label{eq4b}
\end{eqnarray}
Here, $i \tilde{\epsilon}_n(x) = i \epsilon_n + \frac{e}{c} 
\vec{v}_F \cdot \vec{A}(x)$. The initial values for the two quantities
$a(x)$ and $b(x)$ in the bulk superconductor have to be taken as
\begin{eqnarray}
a(-\infty) = \frac{\Delta(-\infty)}{\epsilon_n + \sqrt{\epsilon_n^2 + 
|\Delta(-\infty)|^2}}
\label{eq5a} \\ 
b(+\infty) = \frac{\Delta^\dagger(+\infty)}{\epsilon_n + \sqrt{\epsilon_n^2 + 
|\Delta(+\infty)|^2}}
\label{eq5b}
\end{eqnarray}
Once the order parameter field $\Delta_0 (\vec{r},\Theta)$ and
the vector potential $\vec{A}(\vec{r})$ are known, the initial value
problem for the scalar differential equations (\ref{eq4a}) and (\ref{eq4b})
can be solved by standard numerical techniques (adaptive Runge-Kutta method
for example). For a homogeneous bulk superconductor one confirms that
Eqs. (\ref{eq5a}) and (\ref{eq5b}) indeed fulfil Eqs. (\ref{eq4a}) and 
(\ref{eq4b}). In order to find the local density of states Eq.~(\ref{eq2b})
for a given point $\vec{r}$ and energy $\epsilon$ it is necessary to
solve the Riccati equations (\ref{eq4a}) and (\ref{eq4b}) for a bundle
of trajectories running through the point $\vec{r}$ with different
angles $\Theta$.

It is instructive to see how the Doppler shift method can be obtained from
Eq. (\ref{eq4a}). For that purpose we first decompose the order parameter
$\Delta(x)$ into amplitude and phase via
\begin{equation}
\Delta(x) = \bar{\Delta}(x) e^{i \Phi(x)}
\label{eq6}
\end{equation}
Introducing the function
\begin{equation}
\bar{a}(x) = a(x) e^{-i \Phi(x)}
\label{eq7}
\end{equation}
one arrives at an equation for $\bar{a}$:
\begin{eqnarray}
\lefteqn{
\hbar v_F \frac{\partial}{\partial x} \bar{a}(x) + \left[ 2 \epsilon_n +
2 i \vec{v}_F  \cdot m \vec{v}_s (x)  + \right. } \nonumber \\
& & \hspace*{2.0cm}
\left. + \bar{\Delta}(x) \bar{a}(x) \right] 
\bar{a}(x) - \bar{\Delta}(x) = 0 
\label{eq8}
\end{eqnarray}
where $\vec{v}_s (x) = \frac{1}{2 m} \left( \hbar \vec{\nabla} \Phi (x) - 
\frac{2e}{c} \vec{A}(x) \right)$ is nothing but the gauge invariant 
superfluid velocity
of the supercurrent field distribution. If we now neglect the spacial
derivative $\frac{\partial}{\partial x} \bar{a}(x)$ we find from Eq.
(\ref{eq8})
\begin{equation}
\bar{a}(x) = \frac{\bar{\Delta}(x)}{\bar{\epsilon}_n(x) + 
\sqrt{\bar{\epsilon}_n(x)^2 + \bar{\Delta}(x)^2}}
\label{eq9}
\end{equation}
with $\bar{\epsilon}_n(x) = \epsilon_n + i m
\vec{v}_F  \cdot \vec{v}_s (x)$. Thus, we rediscover the bulk result
apart from a Doppler shift in energy. If we set $\bar{\Delta}(x)$
equal to its homogeneous bulk value using Eqs. (\ref{eq2b}) and (\ref{eq3}) 
we just obtain the usual Doppler shift equation \cite{Schachinger}
\begin{eqnarray}
\lefteqn{
N_{\rm DS}(\vec{r}, \epsilon) = } \label{eq10} \\
& & \frac{1}{2\pi} \int_0^{2 \pi} d \Theta \; 
\mbox{Re} \left\{ \frac{\left| \epsilon - m
\vec{v}_F  \cdot \vec{v}_s (\vec{r}) \right|}
{\sqrt{\left(\epsilon - m
\vec{v}_F  \cdot \vec{v}_s (\vec{r}) \right)^2 - \left| \Delta(\Theta)
\right|^2 }} \right\}
\nonumber
\end{eqnarray}
From this derivation we learn that the Doppler shift method neglects
the gradient term in the Eilenberger equations. This neglection is expected
to be a reasonable approximation as long as the Doppler shift energy
$m \vec{v}_F  \cdot \vec{v}_s (\vec{r})$ is small compared to the
local gap energy $\Delta(\vec{r},\Theta)$. As is well known, this approximation 
fails close to the vortex cores, where the superfluid velocity diverges. 
However, as we will see later, this approximation also fails in the vicinity 
of gap nodes of $\Delta(\vec{r},\Theta)$.

\section{The single phase vortex}
\label{secphase}

In order to make quantitative comparisons between the Doppler shift
method and the Eilenberger approach, we first want to study the
single vortex case. In particular, we restrict ourselves to the
pure 'phase vortex' for which the amplitude of the order
parameter $\Delta(\Theta)$ is assumed to be constant as a function of 
$\vec{r}$ and only its phase is varying. For a $d$-wave superconductor 
taking the magnetic field along the c-axis direction we then have
\begin{equation}
\Delta(\vec{r},\Theta) = \Delta_0 \cos(2\Theta) e^{i\phi}
\label{eq11}
\end{equation}
where $\phi$ denotes the polar angle of $\vec{r}$. We choose
this model for the vortex, because it should be the 'best case'
for the Doppler shift method, since it neglects any spacial
variation of the amplitude of the gap. Such a model for the
vortex is expected to be reasonable for a high-$\kappa$
superconductor at low magnetic fields of the order of the
lower critical field $H_{c1}$. Although this phase vortex does
not possess a core in the usual sense, we will see that there
still exists a core region in the Eilenberger approach, which is
due to the gradient term. One observes that the gradient term
introduces a length scale $\xi= \hbar v_F / \Delta$ into the
problem, even if the amplitude $\Delta$ itself does not vary.
For the single vortex the Doppler shift energy diverges like $r^{-1}$
and we have
\begin{equation}
m \vec{v}_F  \cdot \vec{v}_s (\vec{r}) = \frac{\hbar}{2 r}
\vec{v}_F  \cdot \hat{e}_\phi = \frac{\hbar v_F}{2 r} \sin(\Theta-\phi)
\label{eq12}
\end{equation}
where $\hat{e}_\phi$ is the unit vector in $\phi$-direction.
This equation holds as long as $r$ is smaller than the penetration
depth. Otherwise the screening of the magnetic field has to be taken
into account as well.

We have calculated the local density of states for the Doppler shift
method using Eqs. (\ref{eq10}) and (\ref{eq12}). The local density of
states for the Eilenberger approach is obtained from Eqs. (\ref{eq2b}) 
and (\ref{eq3}) numerically using the Riccati method outlined above.
In order to facilitate the solution a small imaginary part $\delta \le
0.01 \Delta_0$ has been added to the energies on the real axis.

\begin{figure}
  \begin{center}
    \includegraphics[width=0.65\columnwidth,angle=270]{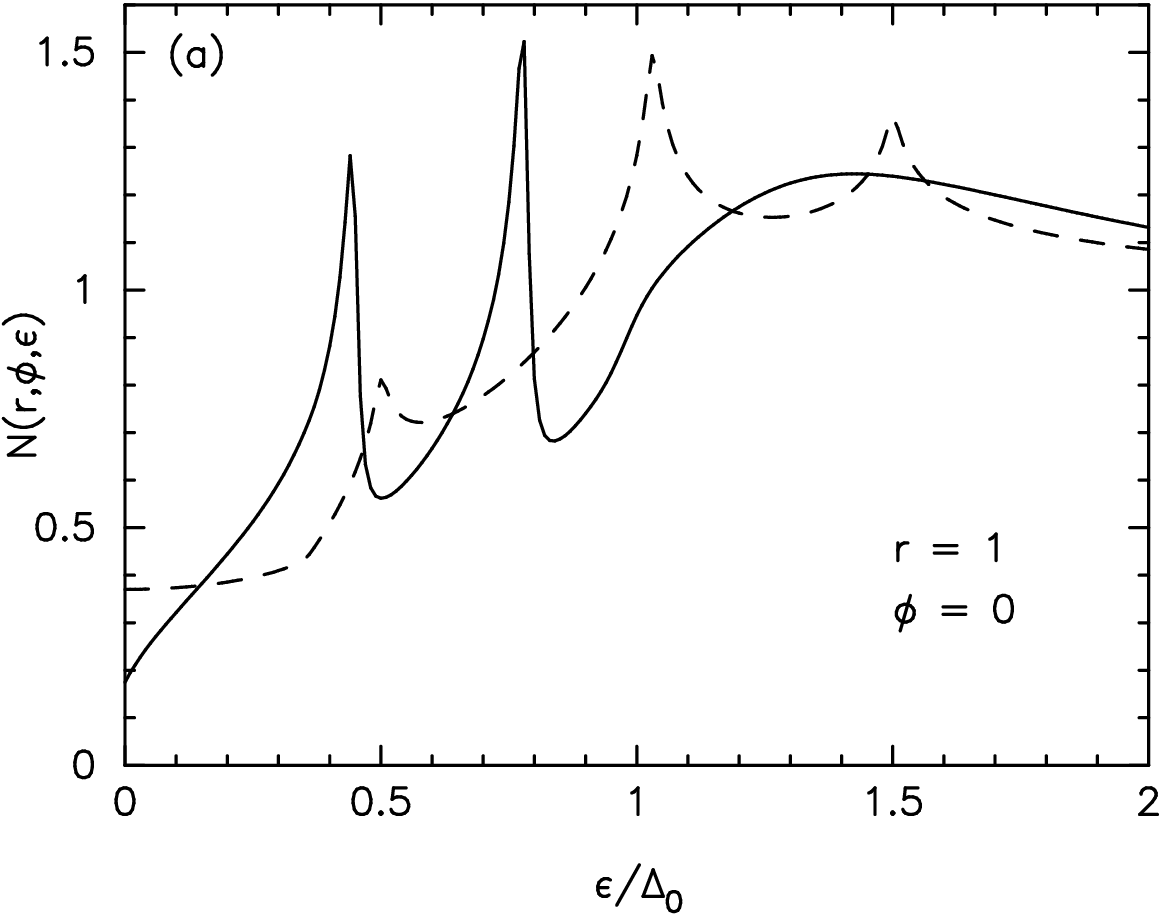}
    \vspace{.4cm}

    \includegraphics[width=0.65\columnwidth,angle=270]{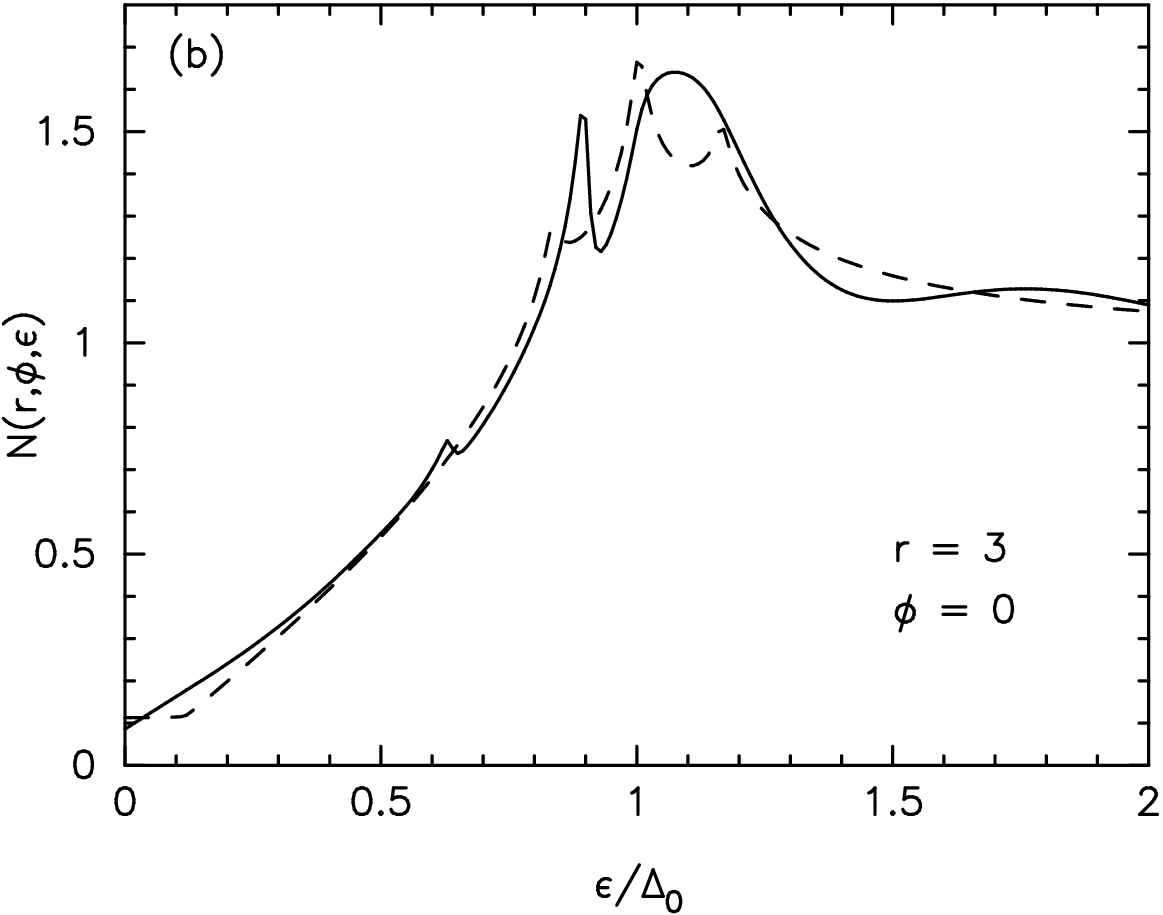}
    \vspace{.2cm}
    \caption{Local density of states in the vicinity of a $d$-wave 
    phase vortex at distances (a) $r=1$ and (b) $r=3$ (in units of 
    the coherence 
    length) from the vortex center in the antinodal direction $\phi=0$.
    The dashed line shows the result from the Doppler shift calculation,
    while the solid line is the result from a numerical solution of the
    Eilenberger equations.
     \label{fig1} }
  \end{center}
\end{figure}

\begin{figure}
  \begin{center}
    \includegraphics[width=0.65\columnwidth,angle=270]{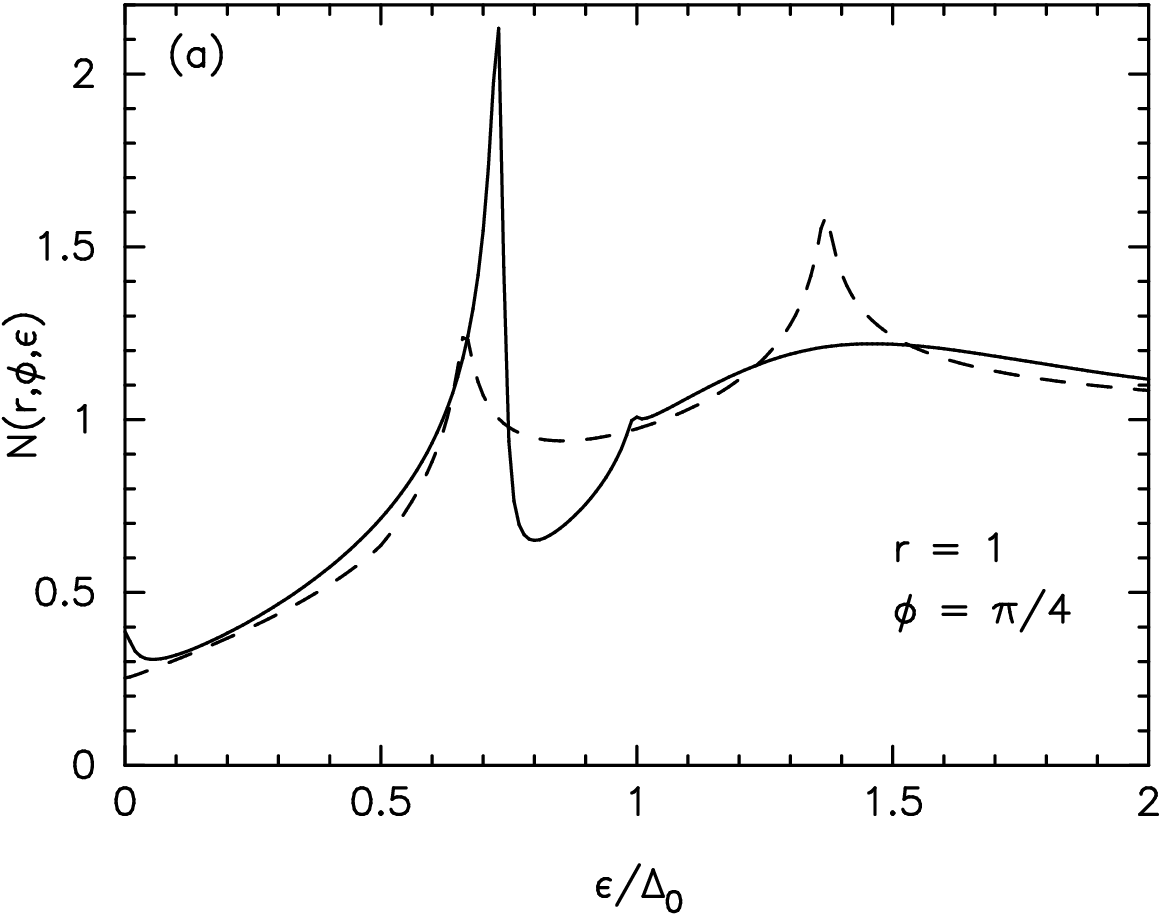}
    \vspace{.4cm}

    \includegraphics[width=0.65\columnwidth,angle=270]{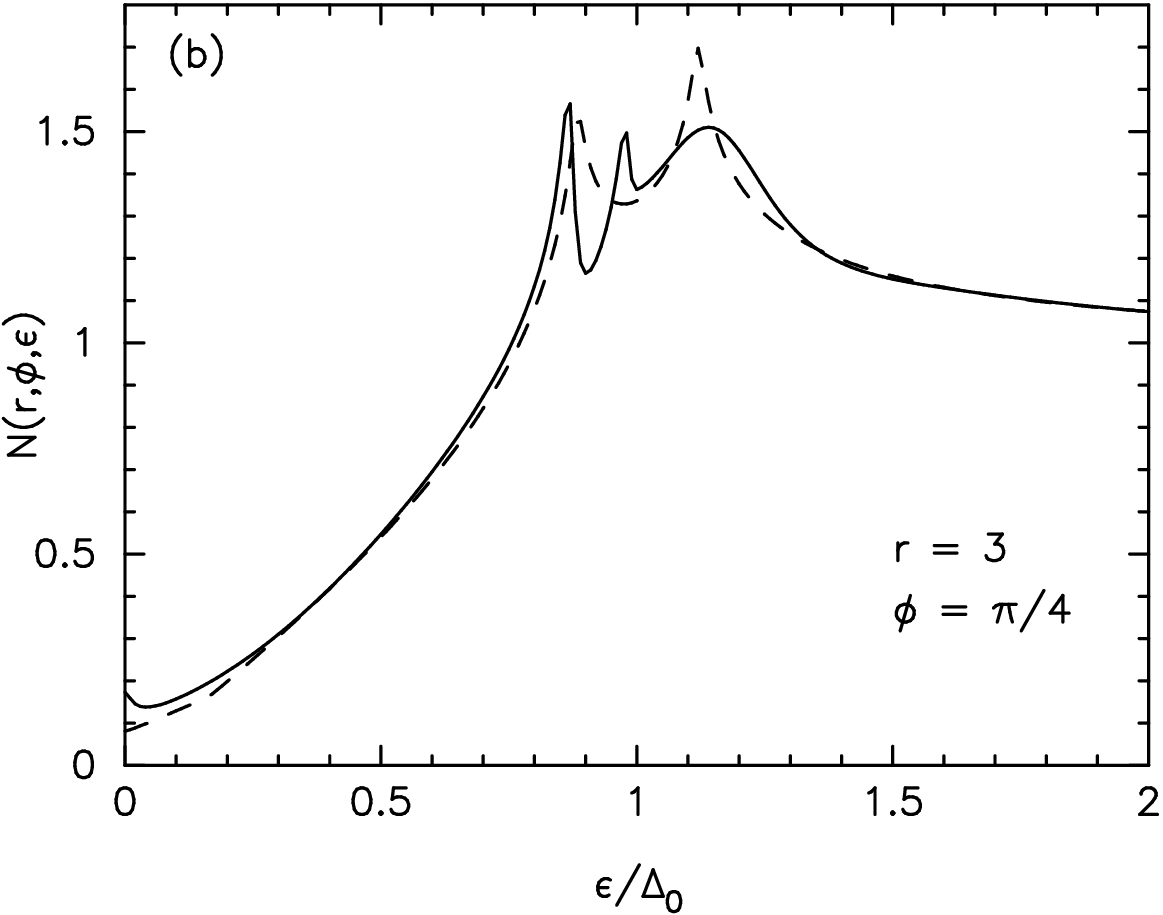}
    \vspace{.2cm}
    \caption{Same as Fig.~\ref{fig1}, but in the nodal direction
    $\phi=\pi/4$ from the vortex center.
     \label{fig2} }
  \end{center}
\end{figure}

In Figs.~\ref{fig1} and \ref{fig2} we are showing our results for
the energy dependence of the local density of states for
a $d$-wave superconducting state. Here and in the following the
density of states is normalized to the normal state value.
In Fig.~\ref{fig1}(a) and (b) the results
in an angular direction of $\phi=0$ are shown for distances $r=1$ and 
$r=3$ from the vortex center (in units of the coherence length $\xi$).
Figs.~\ref{fig2}(a) and (b) show the corresponding results for an
angular direction of $\phi=\pi/4$ (nodal direction).
As becomes clear from these plots, the Doppler shift method
gives resonable results at sufficient high energy and also becomes
better at distances further away from the vortex center. However, at
distances from the vortex of the order of the coherence length
the Doppler shift fails to give the position of the peaks correctly,
which result from core state contributions or scattering
resonances.\cite{Schopohl1,Machida1}  The satellite peaks in the
Doppler-shift method seen in Figs.~\ref{fig1} and \ref{fig2} 
appear at the Doppler shifted gap energy, as has been discussed
recently in Ref. \onlinecite{Schachinger}. A discussion of the
position of the peaks for the $d$-wave vortex within the
quasiclassical approximation can be found in Refs. 
\onlinecite{Schopohl1,Machida1}. 

It has been noted earlier
that in a $d$-wave vortex the core states are not truly localized.
\cite{FT2,Kita} Nevertheless, they give important contributions
to the local density of states, which are not captured by the
Doppler shift method, but are present in the numerical
solution of Eilenberger's equations.
We emphasize that these resonances are still present,
although our phase vortex does not possess a variation of the
magnitude of the gap. Thus, they cannot easily be understood as
'bound states' in the potential well of the gap at the vortex core. 
Instead, these resonances arise due to scattering of the 
quasiparticles at the phase gradient around the vortex center
and thus can be interpreted as Andreev scattering resonances.
For comparison, we also did calculations where we multiplied
Eq.~(\ref{eq11}) by $\tanh (r/\xi_c)$, where $\xi_c$ is an adjustable
parameter independent of $\xi$, which allows us to smoothly cross
over from a vortex possessing a core in the usual sense to our
phase vortex, taking the limit $\xi_c \rightarrow 0$. These calculations
confirm that the core state contributions do not disappear in this
limit, but instead smoothly evolve into the peaks seen in 
Figs.~\ref{fig1} and \ref{fig2}.

Even some distance away from the vortex center, 
the Doppler shift does not give the correct behavior at low
energies, however. As can be seen in Fig.~\ref{fig2}(a) and (b), there is
a small peak at low energies in the local density of states in the
quasiclassical solution. This peak is mainly visible in the
vicinity of the nodal direction $\phi=\pi/4$, as has been noted
before, \cite{Machida1} and is due to
extended core states, which can 'leak out' of the vortex
core region due to the nodes in the gap function. This effect
is also not present in the Doppler shift calculation. 

\begin{figure}
  \begin{center}
    \includegraphics[width=0.65\columnwidth,angle=270]{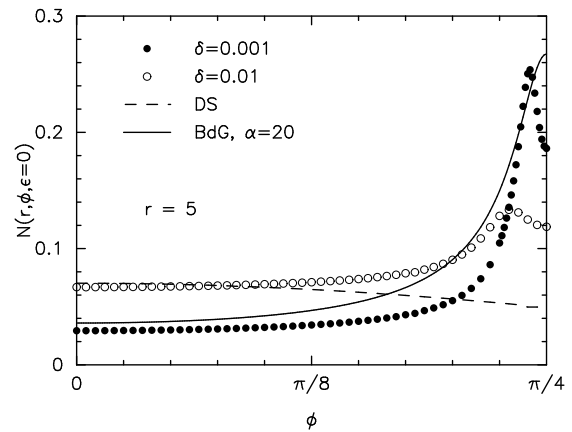}
    \vspace{.2cm}
    \caption{Zero energy local density of states at a distance of $r=5$
     from the vortex center plotted as a function of polar angle $\phi$. 
     The circles are numerical results obtained from the
     solution of Eilenberger's equations using different
     imaginary parts: $\delta=0.001$ (full circles) 
     and  $\delta=0.01$ (open circles). The dashed
     line is the result obtained from the Doppler shift (DS)
     method. The solid line shows the angular dependence
     expected from a solution of the Bogoliubov-de~Gennes (BdG)
     equations linearized around the gap nodes due to
     Mel'nikov.\protect\cite{Melnikov}
     \label{fig3} }
  \end{center}
\end{figure}

This becomes more transparent from Fig.~\ref{fig3}, where we show
the zero energy local density of states at a distance $r=5$ from
the vortex center as a function of polar angle $\phi$. The symbols
show the results from our numerical solution of Eilenberger's
equations, open circles for an imaginary part of $\delta=0.01$ and
solid circles for $\delta=0.001$ (in units of $\Delta_0$). The dashed
line shows the Doppler shift result. (A discussion of the solid line
is found below). From this plot it becomes clear that the solution
of Eilenberger's equations yields increased contributions to the
local density of states at zero energy especially close to the nodal
direction $\phi=\pi/4$, which the Doppler shift method is not able to
capture. We can trace back these increased contributions to
quasiparticle trajectories passing near by the vortex center with a
momentum direction close to the nodal direction $\Theta=\pi/4$ by
looking at the momentum resolved local density of states
$N(r,\phi,\Theta,\epsilon)=\; \mbox{Re} 
\left\{ g(r,\phi,\Theta,\epsilon) \right\}$. On these particular
trajectories the gap is small and the corresponding wavefunction
of the quasiparticle extends very far, 'leaking out' of the
core region. These nonlocal effects
are missed by the Doppler shift method, since it only knows
about the local supercurrent flow. As Fig.~\ref{fig3} shows,
this effect strongly depends on the imaginary part $\delta$ chosen.
It is clear that $\delta$ introduces a finite mean free path into
the problem, which limits the extend to which these quasiparticle
states can contribute to the local density of states far from the
vortex core. Thus, one should expect that this effect is sensitive
to impurity scattering.

One notes that in Fig.~\ref{fig3} the results from Eilenberger's
equations display peaks slightly off the $\phi=\pi/4$ direction.
The position of the peaks depends on $\delta$ and moves closer
to $\pi/4$ when $\delta$ is reduced. The suppression directly at
$\pi/4$ is due to the presence of the $d$-wave gap node in this
direction, because a trajectory passing through the vortex center
in this direction does not 'see' a gap and thus no resonant
Andreev scattering takes place. (For comparison see also Fig. 8d
in Ref. \onlinecite{Machida1}). We observe that the main contribution
to the local density of states in this case is coming from trajectories
slightly off the $\pi/4$ direction, depending on the momentum width
induced by $\delta$.

\begin{figure}
  \begin{center}
    \includegraphics[width=0.65\columnwidth,angle=270]{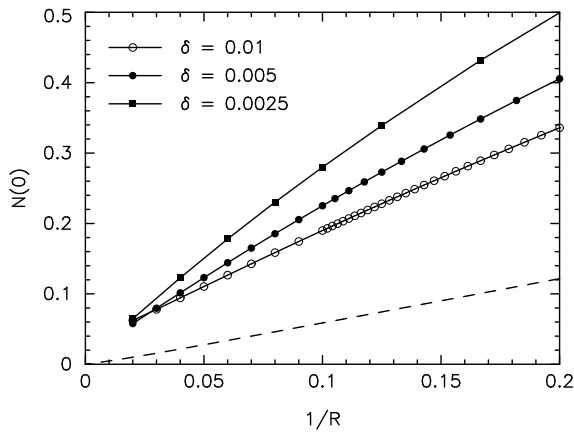}
    \vspace{.2cm}
    \caption{Average density of states within a circle of radius $R$
     (in units of the coherence length $\xi$) 
     around the vortex plotted as a function of $1/R$. The dashed
     line is Volovik's result obtained from the Doppler shift
     method. The symbols are numerical results obtained from a
     solution of Eilenberger's equations using different
     imaginary parts: $\delta=0.01$ (open circles), 
     $\delta=0.005$ (full circles), and  $\delta=0.0025$ (full squares).
     Here, $\delta$ is measured in units of the gap amplitude
     $\Delta_0$. 
     \label{fig4} }
  \end{center}
\end{figure}

This effect also has important consequences for the magnetic field
dependence of the density of states at zero energy. In 
Fig.~\ref{fig4} we show the density of states averaged
over a circle of radius $R$ around the vortex center as a
function of $1/R$, which at low fields and for high $\kappa$
superconductors is proportional to the square-root of the
magnetic field. The Doppler shift method yields a
square-root dependence of the density of states as a function of
magnetic field (a linear variation as a function of $1/R$), 
which has
been noted first by Volovik. \cite{Volovik1} However, the
quasiclassical solution leads to important deviations from
this law. First of all, the slope of these curves is much
higher, and secondly this effect is very sensitive
to the imaginary part chosen in the calculation, as is clear
from the discussion above. While we believe that the field
dependence of these curves will probably not be distinguishable
experimentally from the square-root field dependence, a 
systematic study of the influence of impurity content on
the field dependence of the specific heat at low temperatures
might be able to detect the sensitivity of the slope to
impurity scattering and would be a valuable
confirmation of the presence of these extended core states.

Since we are working here with a quasiclassical approximation one 
might ask, how important quantum effects like the Aharanov-Bohm
effect might be on these extended core states. Indeed, the 
Aharanov-Bohm effect for a $d$-wave vortex has been studied recently
by Mel'nikov within a solution of the Bogoliubov-de~Gennes equations
linearized around the gap nodes. \cite{Melnikov} Within this
quantum mechanical calculation Mel'nikov showed that the angular
dependence of the zero energy density of states far from the
vortex core is determined by the Dirac cone anisotropy
$\alpha=v_F/v_\Delta$, where $v_\Delta$ is the quasiparticle
velocity tangential to the Fermi surface at the node, which is
given by the slope of the gap (in our case $v_\Delta=2 \Delta_0/\hbar k_F$
and thus $\alpha=k_F \xi/2$). For the angular dependence
Mel'nikov found the expression \cite{Melnikov}
\begin{eqnarray}
\lefteqn{
N \left( r, \phi -\frac{\pi}{4}, \epsilon=0 \right) =}
\label{eq12p} \\ & & \frac{4}{\pi}
\frac{\xi}{r} \left\{ \frac{1}{\sqrt{{\alpha^2}
\cos^2 \phi + \sin^2 \phi }} +
\frac{1}{\sqrt{{\alpha^2}
\sin^2 \phi + \cos^2 \phi  }}
 \right\} \nonumber
\end{eqnarray}
(Note, that in Ref.~\onlinecite{Melnikov} the nodal direction
corresponds to $\phi=0$. Therefore, we are shifting this
expression here by $\pi/4$. The prefactor $4/\pi$ can be found from
Mel'nikov's result normalizing it to the normal density of states). 
For comparison this angular
dependence is shown in Fig.~\ref{fig3} for $\alpha=20$ as
the solid line. The angular dependence becomes more pronounced
for higher values of $\alpha$, but estimates from thermal conductivity
measurements on high-$T_c$ cuprates indicate anisotropies in the
range $\alpha \sim 15-20$.\cite{Chiao} Clearly, the contribution from
the extended core states is contained in both the quantum
mechanical and the quasiclassical calculation. Note, that
the full quantum mechanical information about the particle-hole
coherence along the classical trajectories is contained in the
quasiclassical approach (while the quantization perpendicular
to the trajectories is neglected).
Our quasiclassical approach corresponds to taking
the limit $2\alpha=k_F \xi \rightarrow \infty$, but does not
depend on a linearization around the gap node. From the comparison
in Fig.~\ref{fig3} we expect that full quantum interference, i.e.
interference of quasiparticles running classically different ways 
around the vortex line,
will limit (via the Dirac cone anisotropy) the angular dependence
of the density of states. But this comparison also shows that
a very pure sample is needed, if one wants to observe these
quantum mechanical effects, since at shorter mean free paths
the angular dependence is expected to be limited by the impurity
scattering rate.

\section{Vortex lattice}
\label{seclatt}

We now want to focus our attention on the density of states in the
vortex lattice. In the vortex lattice, the order parameter profile
as well as the superfluid flow around the vortices has to be
modified as compared to the single vortex case in Eqs. (\ref{eq11}) 
and (\ref{eq12}). Specifically, here we will use Abrikosov's
wavefunction, which is the exact solution close to the upper
critical field $B_{c2}$.
An arbitrary vortex lattice $\Lambda$ can be spanned by two lattice vectors
$\vec{R}_1=( \omega_1,0 )$ and $\vec{R}_2=( {\rm Re} \; \omega_2, 
{\rm Im} \; \omega_2)$, where we combined the x- and y-component of
$\vec{R}_2$ into the complex quantity $\omega_2$ and we have
chosen the x-axis into the $\vec{R}_1$ direction. Without loss
of generality we may assume $\omega_1>0$ and ${\rm Im} \; \omega_2 >0$.
Using the
quantities $\omega_1$ and $\omega_2$ we can generalize
Abrikosov's wavefunction in the following form:
\begin{eqnarray}
\lefteqn{
\psi_\Lambda \left( x,y \right) = \frac{1}{\cal N} \sum_{n=-\infty}^\infty
\exp \Bigg[ \frac{\pi \left( i x y - y^2 \right)}{\omega_1 {\rm Im} 
\; \omega_2}
 + i \pi n + } \nonumber \\ & & \hspace*{0.5cm} 
+ \frac{i \pi \left( 2n+1 \right) }{\omega_1} \left( x + i y \right) 
+ i \pi \frac{\omega_2}{\omega_1} n \left( n+1 \right) \Bigg]
\label{eq13}
\end{eqnarray}
where ${\cal N}=\left[ \frac{\omega_1}{2 \; {\rm Im} \; \omega_2} 
\exp \left( \pi \frac{{\rm Im} \; \omega_2}{\omega_1} \right) 
\right]^{1/4}$ is a normalization factor such that $|\psi_\Lambda|^2$
averaged over a unit cell $C_\Lambda$ of the vortex lattice becomes unity. 
For a square lattice we have $\omega_2= i \omega_1$, while for
a triangular lattice $\omega_2= \frac{1 + i \sqrt{3}}{2} \omega_1$.
The average magnetic field $\vec{B}_{\rm av}$ in the 
superconductor points into
the z-direction and the gauge has been chosen such that the vector
potential obeys the relation
\begin{equation}
\vec{A} = -\frac{1}{2} \vec{r} \times \vec{B}_{\rm av}
\label{eq14}
\end{equation}
Due to the flux quantization the average magnetic field is
related to the area of the unit cell of the vortex lattice yielding
the relation $B_{\rm av}=\frac{h c}{2 e \omega_1 {\rm Im} \; \omega_2}$.
The superfluid velocity field  $\vec{v}_s (\vec{r})$ for the
Doppler shift calculation can then be 
obtained from Eq.~(\ref{eq13}) using the relation
\begin{eqnarray}
\vec{v}_s (\vec{r}) &=& \frac{\hbar}{4 m i} \frac{\psi_\Lambda^* 
\vec{\nabla} \psi_\Lambda - \psi_\Lambda \vec{\nabla} \psi_\Lambda^* }
{\left| \psi_\Lambda \right|^2} - \frac{e}{m c} \vec{A}
\label{eq15} \\ 
&=& \frac{\hbar}{2 m} \left\{ \frac{{\rm Im} \left[ \psi_\Lambda^* 
\vec{\nabla} \psi_\Lambda \right] }{\left| \psi_\Lambda \right|^2} 
- \frac{\pi}{\omega_1 {\rm Im} \; \omega_2}
\left( \begin{array}{c} -y \\ x \end{array} \right) \right\}
\nonumber 
\end{eqnarray}
Due to the $n^2$ term in the exponent the sum in Eq.~(\ref{eq13}) quickly
converges and only a few terms are sufficient to find 
$\psi_\Lambda \left( x,y \right)$. For our numerical solution of the 
Eilenberger
equations for a $d$-wave superconductor we then use the order parameter
\begin{equation}
\Delta(\vec{r},\Theta) = \Delta_0 \cos(2\Theta) \psi_\Lambda(\vec{r})
\label{eq16}
\end{equation}
When solving the Riccati equations Eqs. (\ref{eq4a}) and 
(\ref{eq4b}) for the vortex lattice we have to ensure that a solution
periodic in the lattice is found. We achieve that by utilizing our
small imaginary part $\delta$ for the energies. For a given point in
space and momentum we choose the starting point of the corresponding
trajectory a long distance of order $1/\delta$ unit cells away.
This gives the solution a distance to relax to the periodic
solution sought, where the relaxation is provided by $\delta$. We
have explicitly checked that the result is periodic and doesn't
change anymore, if we repeat the calculation with a longer
trajectory.

\subsection{An approximate analytical solution}

At high magnetic fields close to the upper critical field $B_{c2}$
it is possible to obtain analytic results for the density of
states averaged over a unit cell of the vortex lattice,
generalizing a method due to Pesch, \cite{Pesch} which has
been used recently also for the study of $d$-wave superconductors.
\cite{Vekhter2} These analytic results
can be obtained as follows. The first step is to replace
the function $g$ on the right hand side in Eq.~(\ref{eq1}) by
its value averaged over a unit cell of the vortex lattice.
This approximation is certainly valid close to the upper critical
field, because there $g$ does not vary strongly within a unit cell.
Then we have
\begin{equation}
L f(\vec{r}, \Theta, 
i \epsilon_n) =  
2 \langle g(\vec{r}, \Theta,i \epsilon_n) \rangle_{C_\Lambda}
\Delta (\vec{r},\Theta)
\label{eq17}
\end{equation}
where $\langle \cdots \rangle_{C_\Lambda}$ denotes an average over
a unit cell of $\vec{r}$ and the operator $L$ is given by
\begin{equation}
L = L \left( \Theta \right) = 2 \left( \epsilon_n - i \frac{e}{c} 
\vec{v}_F \left( \Theta \right) \cdot \vec{A}
\right) + \hbar \vec{v}_F \left( \Theta \right) \cdot \vec{\nabla} 
\label{eq18}
\end{equation}
Assuming without loss of generality that $\epsilon_n>0$ 
Eq.~(\ref{eq17}) can be inverted using the operator identity
\cite{Schopohl2}
\begin{equation}
L^{-1} = \int_0^\infty ds \; e^{-s L}
\label{eq19}
\end{equation}
Since $\langle g \rangle_{C_\Lambda}$ does not depend on the
variable $\vec{r}$ anymore, $L$ only acts upon $\Delta$ and we
find
\begin{equation}
f(\vec{r}, \Theta, i \epsilon_n) =  
2 \langle g(\vec{r}, \Theta,i \epsilon_n) \rangle_{C_\Lambda}
\int_0^\infty ds \; e^{-s L} \Delta (\vec{r},\Theta)
\label{eq20}
\end{equation}
Using the normalization condition Eq.~(\ref{eq2}) we can calculate
the cell average of the square of $g$:
\begin{eqnarray}
\lefteqn{
\langle g^2(\vec{r}, \Theta,i \epsilon_n) \rangle_{C_\Lambda}} \nonumber \\
&=& 1 - \langle f(\vec{r}, \Theta, i \epsilon_n)
f^*(\vec{r}, \Theta + \pi, i \epsilon_n) \rangle_{C_\Lambda}
 \nonumber \\
&=& 1 - \langle g(\vec{r}, \Theta,i \epsilon_n) \rangle_{C_\Lambda}^2
P_\Lambda(\Theta,i \epsilon_n)
\label{eq21}
\end{eqnarray}
where
\begin{eqnarray}
P_\Lambda(\Theta,i \epsilon_n) & =&
4 \int_0^\infty ds_- \int_0^\infty ds_+ \label{eq22} \\
& &
\langle \left( e^{-s_+ L} \Delta (\vec{r},\Theta) \right) 
\left( e^{-s_- L} \Delta^\dagger (\vec{r},\Theta + \pi) \right) \rangle_{C_\Lambda} 
\nonumber 
\end{eqnarray}
Here we have used the relation $g^\dagger(\Theta+\pi)=g(\Theta)$ and
inversion symmetry $L^\dagger(\Theta+\pi)=L(\Theta)$.
Close to the upper critical field we may assume as a second step
that 
\begin{equation}
\langle g^2(\vec{r}, \Theta,i \epsilon_n) \rangle_{C_\Lambda}
=  \langle g(\vec{r}, \Theta,i \epsilon_n) \rangle_{C_\Lambda}^2
\label{eq23}
\end{equation}
which amounts to neglecting spacial fluctuations of $g$. Using this
approximation Eq.~(\ref{eq21}) becomes a closed equation for 
$\langle g \rangle_{C_\Lambda}$ and we finally find
\begin{equation}
\langle g(\vec{r}, \Theta,i \epsilon_n) \rangle_{C_\Lambda}
=  \frac{1}{\sqrt{1+P_\Lambda(\Theta,i \epsilon_n) }}
\label{eq24}
\end{equation}
As it turns out, Eq.~(\ref{eq22}) can be evaluated analytically for
the Abrikosov vortex lattice Eq.~(\ref{eq13}). After some algebra 
\cite{Schopohl2} we find
\begin{equation}
P_\Lambda(\Theta,i \epsilon_n) = \frac{ 2 \Delta_0^2 \cos^2(2\Theta)}
{\epsilon_n^2 } z^2
\left[ 1- \sqrt{\pi} z \; w \left(i z \right) \right]
\label{eq25}
\end{equation}
where 
\begin{equation}
z=\frac{\sqrt{2\omega_1 {\rm Im} \; \omega_2} 
 \; \epsilon_n }{\sqrt{\pi} v_F \hbar}
\label{eq26}
\end{equation}
Here, the function $w$ is Dawson's integral and is related to
the complement of the Error function via
\begin{equation}
w(iz)= \frac{1}{i \pi} \int_{-\infty}^{\infty} 
\frac{e^{-t^2} dt}{t-iz} = e^{z^2} {\rm erfc}(z)
\label{eq27}
\end{equation}
Some properties of this function can be found in Ref. \onlinecite{Abram}.
We can make these equations a little bit more transparent, if we introduce
two length scales: the coherence length $\xi=v_F \hbar / \Delta_0$ and
$a_\Lambda^2=\omega_1 {\rm Im} \; \omega_2 = \Phi_0/B_{\rm av}$ with 
$\Phi_0$ being the flux quantum. $a_\Lambda^2$ is the area of a unit cell of 
the vortex lattice and for a square lattice $a_\Lambda$ is just the distance
between neighboring vortices. We remark that also the coherence length
$\xi$ depends on magnetic field, since $\Delta_0$ has to be determined
from the gap equation in the presence of the field. 
Expressed in these quantities we have
\begin{equation}
z=\sqrt{\frac{2}{\pi}} \frac{a_\Lambda}{\xi} \frac{\epsilon_n }{\Delta_0}
\label{eq28}
\end{equation}
To obtain the cell average of the density of states, Eq.~(\ref{eq24})
has to be integrated over the angle $\Theta$ and analytically continued
to real frequencies $i\epsilon_n \rightarrow \epsilon + i 0^+$. Since
$P_\Lambda$ only depends on $\Theta$ via the $\cos^2(2\Theta)$ term,
the angular integral is just a complete elliptical integral and we find
the analytical result
\begin{eqnarray}
N(\epsilon) &=& \frac{1}{2\pi} \int_0^{2 \pi} d \Theta \; \mbox{Re} 
\left\langle g(\vec{r}, \Theta,i \epsilon_n \rightarrow \epsilon + i 0^+) 
\right\rangle_{C_\Lambda} \nonumber \\
&=& \frac{2}{\pi} \mbox{Re} \left\{ K \left( k \right) \right\}
\label{eq29}
\end{eqnarray}
with
\begin{equation}
k^2 =  - \frac{4}{\pi} \frac{a_\Lambda^2}{\xi^2}
\left[ 1+ i\sqrt{2} \frac{a_\Lambda}{\xi} \frac{\epsilon }{\Delta_0}
\; w \left( \sqrt{\frac{2}{\pi}} \frac{a_\Lambda}{\xi} 
\frac{\epsilon }{\Delta_0} \right) \right]
\label{eq30}
\end{equation}
For comparison, for an $s$-wave superconductor the $\cos^2(2\Theta)$ factor
in Eq.~(\ref{eq25}) has to be dropped and we just find
\begin{equation}
N(\epsilon) =\mbox{Re} \left\{ \frac{1}{\sqrt{1-k^2}} \right\}
\label{eq31}
\end{equation}
It is instructive to consider the low and high field limits of these
expressions. At high magnetic field $B_{c2}$ the coherence length
$\xi$ diverges. In this limit $k \rightarrow 0$ and
we find the normal state result $N(\epsilon) \rightarrow 1$ for both 
$s$- and $d$-wave superconductor as one should expect. For low
magnetic fields $\xi$ becomes constant, but $a_\Lambda$ diverges. In 
this limit one can use the asymptotic expansion of Dawson's integral
\cite{Abram}
\begin{equation}
1- \sqrt{\pi} z \; w \left(i z \right) \sim \frac{1}{2 z^2}
\label{eq32}
\end{equation}
Then we have $k^2 \sim \Delta_0^2/\epsilon^2$. Inserting this
into Eqs.~(\ref{eq29}) and (\ref{eq31}) we just rediscover the
zero field density of states of the $d$- and $s$-wave superconductor,
respectively. It is interesting to note that this high field approximation
also appears to give the correct zero field limit.

\begin{figure}
  \begin{center}
    \includegraphics[width=0.65\columnwidth,angle=270]{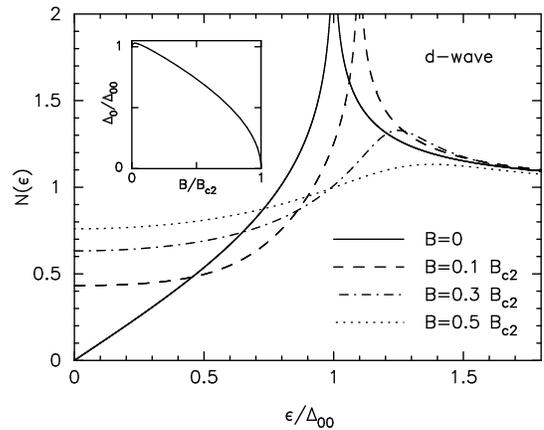}
    \vspace{.2cm}
    \caption{Average density of states in the vortex state of a
     $d$-wave superconductor calculated using the analytical 
     formula Eq.~(\ref{eq29}) for different field strengths:
     $B_{\rm av}=0$ (solid line), $B_{\rm av}=0.1 B_{c2}$
     (dashed line), $B_{\rm av}=0.3 B_{c2}$ (dashed-dotted line),
     and $B_{\rm av}=0.5 B_{c2}$ (dotted line). The energy is
     normalized to the gap value at zero field and temperature
     $\Delta_{00}=\Delta_{0}(T=0,B=0)$. The inset shows the field
     dependence of the zero temperature gap $\Delta_{0}/\Delta_{00}$.
     \label{fig5} }
  \end{center}
\end{figure} 

In Fig.~\ref{fig5} we show the density of states calculated from
Eq.~(\ref{eq29}) for different magnetic fields at zero temperature.
In order to find the appropriate value of $a_\Lambda/\xi$ for a given field
and temperature, it is necessary to solve the gap equation in order
to find $\Delta_0(T,B_{\rm av})$. Using the approximation Eq.~(\ref{eq20})
the gap equation for the $d$-wave case can be cast into the form
\begin{eqnarray}
\lefteqn{
\ln \left( \frac{T}{T_c} \right) =}
\label{eq32p} \\ & & 
T \sum_{\epsilon_n} \int_0^{2\pi} d\Theta
\frac{2 \cos^2 ( 2 \Theta)}{\epsilon_n}
\left\{
\frac{\sqrt{\pi} z \; w \left( iz \right)}{\sqrt{1 + 
P_\Lambda(\Theta,i \epsilon_n)}} -1
\right\} \nonumber
\end{eqnarray}
with $z$ depending on $\epsilon_n$ and $B_{\rm av}$ via Eq.~(\ref{eq26}).
The angular integration over $\Theta$ can even be further reduced
to complete elliptical integrals, if desired. The field dependence
of the gap at zero temperature obtained from this equation is shown in 
the inset to Fig.~\ref{fig5}. The energy scale in Fig.~\ref{fig5}
has been normalized to the zero field and temperature value of the
gap $\Delta_{00}=\Delta_0(T=0,B_{\rm av}=0)$. 
In the inset it can be noted that the gap $\Delta_0$ as a function
of magnetic field rises above $\Delta_{00}$ at low field before
reaching its zero field value $\Delta_{00}$ at $B_{\rm av}=0$.
This is a property of Eq.~(\ref{eq32p}) and we attribute it to the fact
that the approximation used in the present section is valid only at
high field and at zero field. Fig.~\ref{fig5} shows that,
although the gap closes as a function of field $B_{\rm av}$, the peaks
in the density of states move upwards in energy. This means that
in the vortex state the peak-to-peak distance in the density
of states is not an appropriate measure of the energy gap
$\Delta_0$ anymore. This fact is important to realize in 
interpretation of tunneling spectra in the vortex state, for example.

\subsection{The approximation due to Brandt, Pesch, and Tewordt}

In the literature there exists another approximation for
the average density of states in type~II superconductors at high
magnetic fields, which is due to Brandt, Pesch, and Tewordt \cite{BPT}
and also has been used recently to study unconventional
superconductors. \cite{WonMaki2,FayTew}
In their original paper these authors derived their method from
Gorkov's equation. \cite{Delrieu} We discovered a method to re-derive their results
from the Eilenberger equation using the Riccati equations (\ref{eq4a})
and (\ref{eq4b}), which we want to briefly sketch here.

Instead of replacing the function $g$ by its average value in 
Eq.~(\ref{eq1}) we can use an approximation similar in spirit in
the Riccati equations (\ref{eq4a}) and (\ref{eq4b}), linearizing
them by replacing the terms $\Delta^\dagger a$ and
$\Delta b$ by their averages over a unit cell. For the
function $a$ one obtains the equation
\begin{equation}
\bar{L} a(\vec{r}, \Theta, 
i \epsilon_n) =  
\Delta (\vec{r},\Theta)
\label{eq33}
\end{equation}
where $\bar{L}$ is just the operator in Eq.~(\ref{eq18}), except that
$\epsilon_n$ now has to be replaced by 
\begin{equation}
\bar{\epsilon}_n = \epsilon_n + \frac{1}{2}
\langle  \Delta^\dagger a \rangle_{C_\Lambda}
\label{eq34}
\end{equation}
Eq.~(\ref{eq33}) can be inverted the same way as above and after
some algebra we obtain
for the average $\langle \Delta^\dagger a \rangle_{C_\Lambda}$:
\begin{eqnarray}
\lefteqn{
\langle  \Delta^\dagger a \rangle_{C_\Lambda} =
\int_0^\infty ds \; \langle \Delta^\dagger (\vec{r},\Theta) e^{-s \bar{L}} 
\Delta (\vec{r},\Theta) \rangle_{C_\Lambda} }\nonumber \\
&=&  \Delta_0 \cos^2(2\Theta) \frac{a_\Lambda}{\sqrt{2} \xi} w(i \bar{z}) 
= 2 \left( \bar{\epsilon}_n - \epsilon_n \right)
\label{eq35}
\end{eqnarray}
where now
\begin{equation}
\bar{z}=\frac{\sqrt{2\omega_1 {\rm Im} \; \omega_2} 
 \; \bar{\epsilon}_n }{\sqrt{\pi} v_F \hbar}
=\sqrt{\frac{2}{\pi}} \frac{a_\Lambda}{\xi} \frac{\bar{\epsilon}_n }{\Delta_0}
\label{eq36}
\end{equation}
Eq.~(\ref{eq35}) is an implicit equation for $\bar{\epsilon}_n$
or equivalently for $\bar{z}$ and can be written in the form
\begin{equation}
i \bar{z}=\sqrt{\frac{2}{\pi}} \frac{a_\Lambda}{\xi} 
\frac{i \epsilon_n }{\Delta_0} + i \cos^2(2\Theta) 
\frac{a_\Lambda^2}{2 \sqrt{\pi} \xi^2} w(i \bar{z})
\label{eq37}
\end{equation}
From the inversion of Eq.~(\ref{eq33}) and the equivalent one for the
function $b$ one can also obtain the cell average of the product
$ab$:
\begin{equation}
\langle  a b \rangle_{C_\Lambda} =
\cos^2(2\Theta) \frac{a_\Lambda^2}{\pi \xi^2} 
\left[ 1- \sqrt{\pi} \bar{z} \; w \left(i \bar{z} \right) \right]
\label{eq38}
\end{equation}
Using this result we can obtain an approximation for the cell
average of the propagator $g$ using Eq.~(\ref{eq3})
\begin{eqnarray}
\lefteqn{
\langle g(\vec{r}, \Theta,i \epsilon_n) \rangle_{C_\Lambda}
\approx \frac{2}{1+\langle  a b \rangle_{C_\Lambda}} - 1}
\nonumber \\
& = & \frac{2}{1+\cos^2(2\Theta) \frac{a_\Lambda^2}{\pi \xi^2} 
\left[ 1- \sqrt{\pi} \bar{z} \; w \left(i \bar{z} \right) \right]
} - 1
\label{eq39}
\end{eqnarray}
where $\bar{z}$ has to be obtained from a solution of Eq.~(\ref{eq37})
for each $\epsilon_n$ and $\Theta$. The set of equations (\ref{eq37})
and (\ref{eq39}) just corresponds to Eqs.~(18) and (16) in Ref.
\onlinecite{BPT}. (The parameter $\Lambda$ in that work corresponds 
to $a_\Lambda/\sqrt{2 \pi}$ in our case).

This derivation shows that the approximation due to Brandt, Pesch, 
and Tewordt can also be justified from quasiclassical
Eilenberger theory. However, in contrast to the approximate
analytical method derived in the previous subsection, here
one has to solve a nonlinear implicit equation, which
makes the calculation of the density of states more
difficult. An expansion of Eqs.~(\ref{eq39}) and (\ref{eq24})
in terms of $\Delta_0$ shows that both methods give the same
result up to order $|\Delta_0|^3$.

\subsection{Comparison with numerical results}

We now want to present a comparison of the four methods outlined
above: the Doppler shift method (DS), the approximate analytical
solution (AA), the method due to Brandt, Pesch, and Tewordt (BPT),
and a full numerical solution of Eilenberger's equations (EE)
using the Riccati equations. In all four cases we will base
our calculations on an Abrikosov square lattice. However,
the results do not differ very much, if a triangular lattice
is used. In fact, the density of states within the methods AA 
and BPT does not depend on the lattice structure, as is clear
from the derivation above.

In the following we will compare our results using the
parameter $\xi/a_\Lambda$. This parameter is a measure of the
average magnetic field $B_{\rm av}$ inside the superconductor
and at low field strengths $\xi/a_\Lambda \propto \sqrt{B_{\rm av}}$.
At higher field strengths its field dependence has to be
determined from a solution of the gap equation, because
$\xi$ depends on $\Delta_0$, which decreases with increasing
field. From such a solution we have determined that
$\xi/a_\Lambda=1$ corresponds to about $B_{\rm av} \approx 0.5 B_{c2}$
and $\xi/a_\Lambda=0.2$ corresponds to about $B_{\rm av} \approx 0.04 B_{c2}$
at zero temperature.

\begin{figure}
  \begin{center}
    \includegraphics[width=0.65\columnwidth,angle=270]{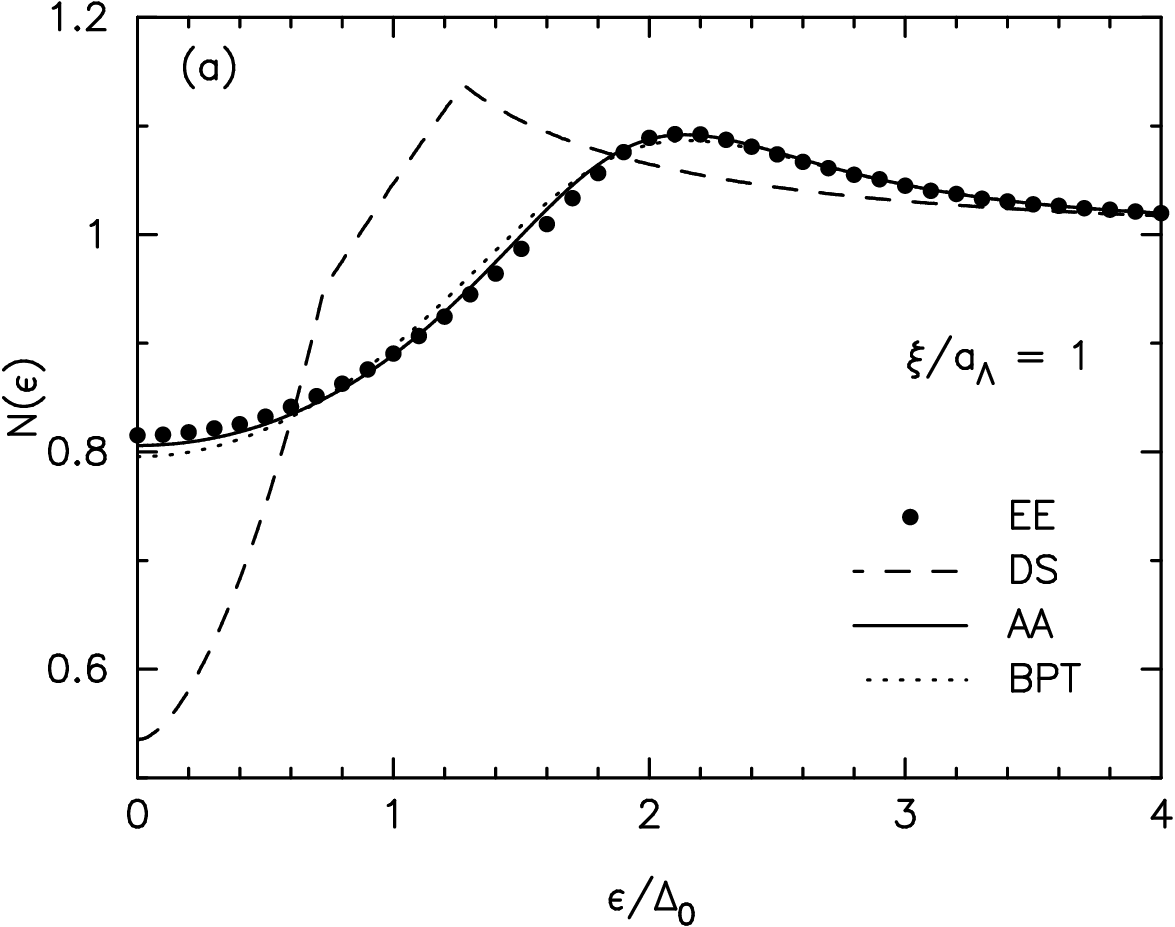}
    \vspace{.4cm}

    \includegraphics[width=0.65\columnwidth,angle=270]{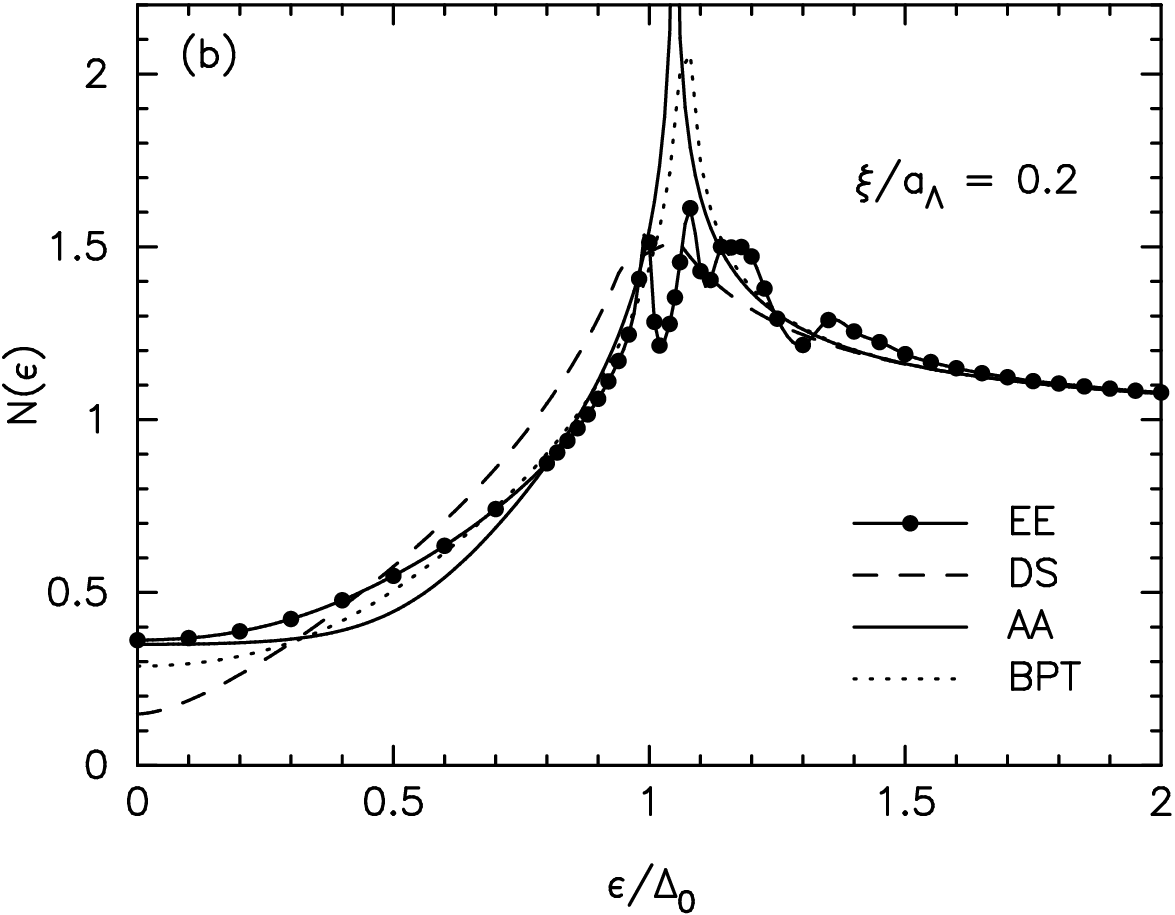}
    \vspace{.2cm}
    \caption{Energy dependence of the density of states averaged
     over a unit cell of the vortex lattice for (a) $\xi/a_\Lambda=1$ 
     and (b) $\xi/a_\Lambda=0.2$, corresponding to magnetic fields
     of about $0.5 B_{c2}$ and $0.04 B_{c2}$, respectively. The dots
     are numerical solutions of the Eilenberger equations, the dashed
     line shows the Doppler shift result, the dotted line the result
     from the method due to Brandt et al, and the solid line the
     approximate analytical result described in the text.
     \label{fig6} }
  \end{center}
\end{figure} 

In Fig.~\ref{fig6} we show the cell average of the density of
states as a function of energy for the four methods. In 
Fig.~\ref{fig6}(a) the comparison for $\xi/a_\Lambda=1$ is shown, 
Fig.~\ref{fig6}(b) shows the results for $\xi/a_\Lambda=0.2$. At high 
magnetic field ($\xi/a_\Lambda=1$) the numerical solution of the 
Eilenberger equations
(EE, black dots), and the methods AA (solid line) and BPT (dotted 
line) are all in very close agreement with each other, the
approximate analytical (AA) result being somewhat closer to the
numerical solution. The Doppler shift method (dashed line), however, 
strongly deviates from these results. This result is not surprising, 
because the Doppler shift method neglects any contributions from vortex
core scattering, while the methods AA and BPT are expansions around
$B_{c2}$ and are expected to give better results at high fields.

Fig.~\ref{fig6}(b) shows the comparison at a considerably
smaller field ($\xi/a_\Lambda=0.2$). In this field range the four methods
yield different results, especially close to the gap edge. The numerical
result clearly shows Tomasch resonances, \cite{Tomasch} which the other 
three methods are not able to get. At high energy all methods converge
to each other. At low energies, which are especially important for
the thermodynamics of the system, the Doppler shift method gives the
strongest deviation, while the AA method is closest to the numerical
solution.

\begin{figure}
  \begin{center}
    \includegraphics[width=0.65\columnwidth,angle=270]{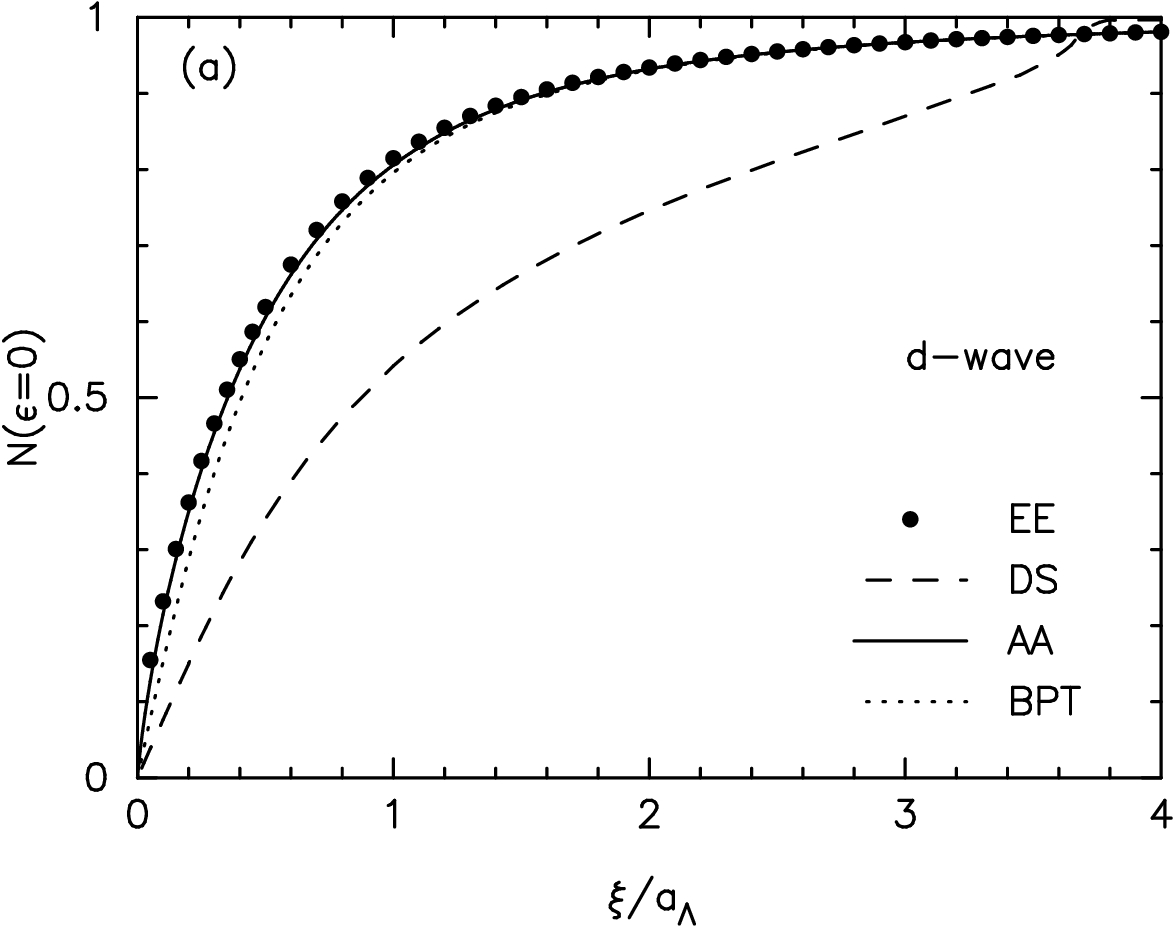}
    \vspace{.4cm}

    \includegraphics[width=0.65\columnwidth,angle=270]{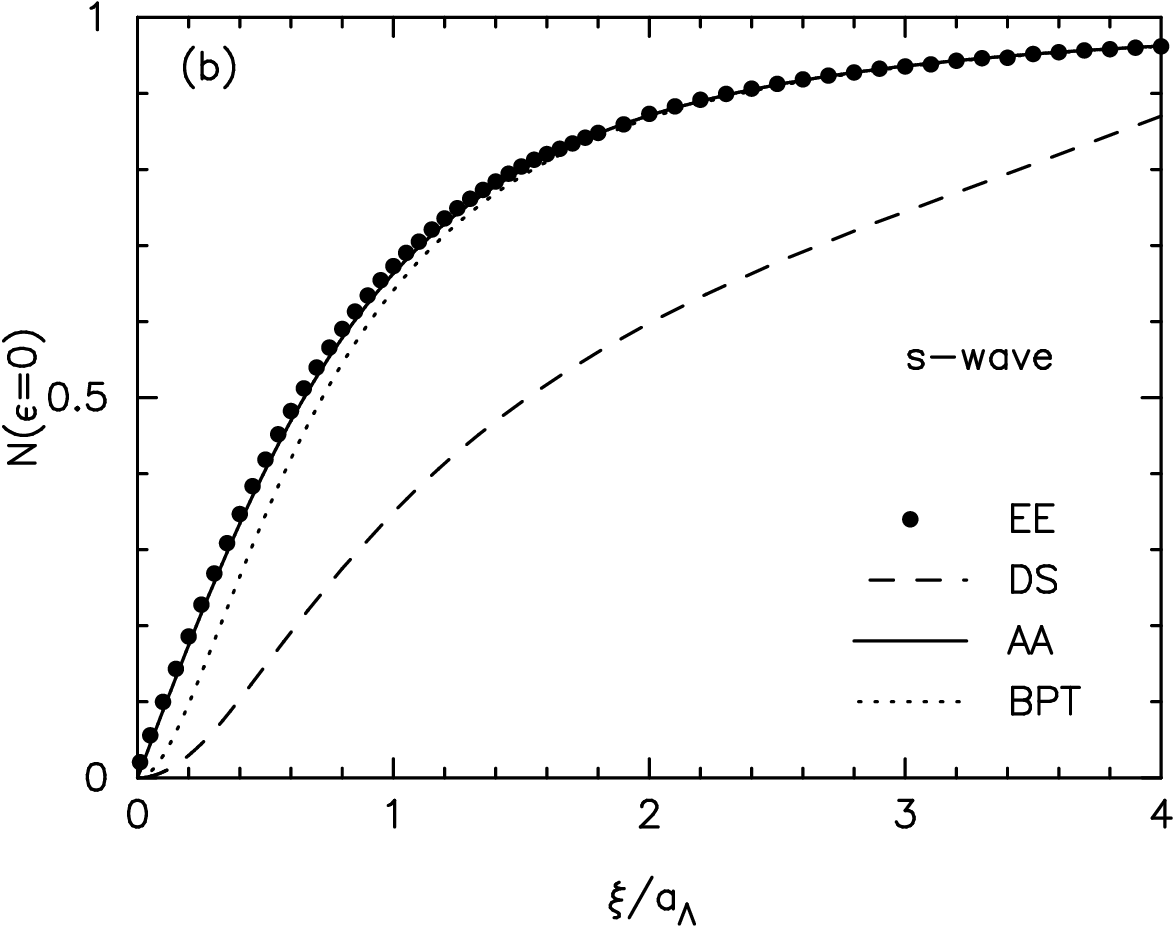}
    \vspace{.2cm}
    \caption{Zero energy density of states as a function of the
    parameter $\xi/a_\Lambda$ for an (a) $d$-wave and an (b) $s$-wave 
    superconductor. The four methods are denoted by the
    same line patterns as in Fig.~\ref{fig6}.
     \label{fig7} }
  \end{center}
\end{figure} 

In Fig.~\ref{fig7} we show the density of states at zero energy as
a function of the parameter $\xi/a_\Lambda$ for both $d$- and $s$-wave
superconductor. From these plots it can be seen that in this
zero energy limit the AA method appears to be astonishingly close 
to the numerical results over the whole field range. At low field
the Doppler shift method gives a density of states proportional
to $\xi/a_\Lambda$ for the $d$-wave case, which corresponds to Volovik's
$\sqrt{B_{\rm av}}$ law. As seen already in the single vortex case
above, the size of the slope turns out to be considerably smaller 
than in the numerical solution, however. For the $s$-wave case
the Doppler shift method and the BPT method both yield a linear
$B_{\rm av}$ field dependence at low fields, while the numerical 
solution and the AA method give a $\sqrt{B_{\rm av}}$ behavior as 
well. We want to caution, however, that in this field range the
Abrikosov vortex lattice Eq.~(\ref{eq13}), which we were using here,
is not an appropriate groundstate wavefunction anymore and 
important corrections from higher Landau levels are expected.
\cite{EHBrandt,EHBrandt2}
In this field range a fully selfconsistent calculation of the
vortex lattice becomes necessary, \cite{Machida1,Machida2} which will 
lead to further corrections, like for example the Kramer-Pesch effect 
\cite{KramerPesch} and the vortex core shrinking effect. \cite{Machida2}

\section{Conclusions}
\label{seccon}

We made a detailed comparison of a numerical solution of
the quasiclassical Eilenberger equations for $s$- and $d$-wave
superconductors in the vortex state with several approximate 
methods. We studied
both the single vortex case and the vortex lattice within
a magnetic field directed along the c-axis.
For the single vortex we found that the Doppler shift method
for a $d$-wave gap not only fails in the vortex core region, 
but also misses important contributions from core states
extending into the nodal directions far away from the vortex
core. These contributions give important corrections to the
density of states especially at low energies, and thus affect
the thermodynamics of the system. In particular, corrections
to Volovik's law are found, which are expected to be very
sensitive to impurity scattering and should be observable
via systematic impurity studies of the field dependence of the 
specific heat. We expect quantum mechanical effects like
the Aharanov-Bohm effect to become visible only in very clean
samples.

In the vortex lattice there are other approximate methods,
which are preferred over the Doppler shift method. Here,
we studied two methods which are valid near the upper
critical field $B_{c2}$: the method due to Brandt, Pesch,
and Tewordt and an approximate analytical method, 
generalizing a method due to Pesch. We showed how the
method due to Brandt, Pesch, and Tewordt can be derived
from the Eilenberger equations using the Riccati equations.
At low fields both of these methods are not able to get
the Tomasch resonances, which are present in the numerical
solution of the Eilenberger equations. However, especially
the approximate analytical method is impressively close
to the numerical results at low energies over the whole
field range. Since this method is also more convenient
than the method due to Brandt, Pesch, and Tewordt,
avoiding a solution of an implicit equation and giving
analytical results in the clean limit, we recommend the
use of this method in the high field range.

\acknowledgments

We would like to thank E.~H.~Brandt, R.~P.~H\"ubener,
K.~Maki, A.~S.~Mel'nikov, P.~Miranovic, L.~Tewordt, 
and C.~C.~Tsuei for valuable discussions. Thanks are also due to
P.~J.~Hirschfeld for bringing Ref.~\onlinecite{Vekhter2} to 
our attention.

\end{document}